\documentstyle[epsfig,12pt]{article}
\begin{document}
\hspace*{1cm}
{\Large\bf Exclusive Nonleptonic Decays of Heavy

\hspace*{1cm}
Baryons in a Relativistic Quark Model}\\[2mm]
\hspace*{1cm}
{\it
     \underline{V.\ E.\ Lyubovitskij$^{a,c}$},
     M.\ A.\ Ivanov,$^a$
     J.\ G.\ K\"{o}rner,$^b$
     A.\ G.\ Rusetsky$^{a,d}$\\[2mm]
\hspace*{1cm}
{\small $^a$JINR, Dubna, $^b$Mainz University,
$^c$Tomsk University, $^d$Tbilisi University}\\[2mm]

\abstract

\noindent
Exclusive nonleptonic decays of bottom and charm baryons are studied
within a relativistic quark model. We include
factorizing as well as nonfactorizing contributions to the decay
amplitudes.

\vspace*{.3cm}


\vspace*{.3cm}
In the near future one can expect large quantities of new data on exclusive
charm and bottom baryon nonleptonic decays which calls for a comprehensive
theoretical analysis of these decays. There exist a number of theoretical
analysis of exclusive nonleptonic heavy baryon decays in the
literature$^{1-6}$ including predictions for their angular
decay distributions. The analysis of nonleptonic baryon decays is complicated by the necessity
of having to include nonfactorizing contributions. In Fig.1 the Feynman
diagrams which contribute to the amplitude of nonleptonic decays of heavy
baryons are given: factorizing diagram (I), nonfactorizing diagrams
(IIa, IIb, III).  One thus has to go
beyond the factorization approximation which had proved quite useful in
the analysis of the exclusive nonleptonic decays of heavy mesons. There
have been some theoretical attempts to analyse nonleptonic heavy baryon
decays using factorizing contributions alone$^5$, the argument
being that W-exchange contributions can be neglected in analogy to the
power suppressed W-exchange contributions in the inclusive nonleptonic
decays of heavy baryons. One might even be tempted to drop the
nonfactorizing contributions on account of the fact that they are
superficially proportional to $1/N_c$. However, since $N_c$-baryons
contain $N_c$ quarks an extra combinatorial factor proportional to $N_c$
appears in the amplitudes which cancels the explicit diagrammatic $1/N_c$
factor$^{1,2}$. There is now ample empirical evidences in
the $c\to s$ sectors that nonfactorizing diagrams cannot be neglected.
For example, in the charm sector the two observed decays
$\Lambda_c^+\to \Xi^0 K^+$ and $\Lambda_c^+ \to \Sigma \pi$ can only
proceed via nonfactorizing diagrams. Their sizeable observed branching
ratios may thus serve to obtain a measure of the size of the nonfactorizing
contributions.

In this paper we use a relativistic quark model developed
in Refs.$^7$ to calculate nonleptonic decays of heavy baryons.
This model has been successfully applied to the description of the
electromagnetic properties of nucleons$^{8}$ and has been extended
to an analysis of the semileptonic decays of heavy baryons$^{9}$.
The detailed description of the model can be found in Refs.$^{7-9}$.
Since the evaluation of the nonleptonic decay amplitudes involves
three-loop diagrams with nonlocal vertices (see Fig.1), we shall make some
simplifying assumptions. These assumptions, on the one hand, have a clear
physics motivation (we reproduce the analysis$^{1}$ of the spin structure
of decay amplitudes in the spectator quark model$^{10,11}$), and, on the
other hand, allow one to evaluate both the factorizing and the nonfactorizing
contributions to nonleptonic baryon decays. This can be achieved by assigning
the projector $V_+=(1+\not\! v)/2$ to each light quark field in the
baryon-quark vertex, and by using the static approximation for
$u$, $d$ and $s$ quark propagators in momentum space $S_q(k)=1/m_q$
with $m_q$ being a static quark mass. See, further details
in Ref.$^6$

Below we present our numerical results for decay rates and
asymmetry parameters. Model parameters are determined from
a fit to known branching ratios of nonleptonic decays
$\Lambda^+_c\to\Lambda^0\pi^+$, $\Lambda^+_c\to \Sigma^0\pi^+$,
$\Lambda^+_c\to \Sigma^+\pi^0$, $\Lambda^+_c\to p\bar K^0$
and $\Lambda^+_c\to \Xi^0 K^+$ (see, Ref.$^6$).
The masses of hadrons are taken from Ref.$^{12}$

In Tables 1  and 2 we give the results for the decay rates and asymmetry
parameters for the heavy-to-light and $b\to c\bar u d$ nonleptonic decays,
respectively.

\vspace*{.5cm}
{\bf Table 1.} Branching ratios (in $\%$) for heavy-light transitions.
\vspace*{.2cm}
\begin{center}
\def\arraystretch{1}
\begin{tabular}{|c|c|c|c|c|c|}
\hline\hline
Process & K\"{o}rner,             & Xu,
& Cheng,             & \,\,\,\,\, Our \,\,\,\,\, & Experiment$^{12}$ \\
        & Kr\"{a}mer $^1$ & Kamal$^3$
& Tseng$^4$ & & \\
\hline
$\Lambda^+_c\to \Lambda \pi^+$ & 0.76 & 1.67 & 0.91 & 0.79 & 0.79$\pm$ 0.18\\
\hline
$\Lambda^+_c\to \Sigma^0 \pi^+$ & 0.33 & 0.35 & 0.74 & 0.88 & 0.88$\pm$ 0.20\\
\hline
$\Lambda^+_c\to \Sigma^+ \pi^0$ & 0.33 & 0.35 & 0.74 & 0.88 & 0.88$\pm$ 0.22\\
\hline
$\Lambda^+_c\to \Sigma^+ \eta$ & 0.16 & & & 0.11 & 0.48$\pm$ 0.17\\
\hline
$\Lambda^+_c\to \Sigma^+ \eta^\prime$ & 1.28 & & & 0.12& \\
\hline
$\Lambda^+_c\to p \bar K^0$ & 2.16 & 1.24 & 1.30 & 2.06 & 2.2$\pm$ 0.4\\
\hline
$\Lambda^+_c\to \Xi^0 K^+$ & 0.27 & 0.10 & & 0.31 &  0.34$\pm$ 0.09\\
\hline
$\Xi^+_c\to \Sigma^+ \bar K^0$ & 5.11 & 0.35 & 0.67 & 3.08 & \\
\hline
$\Xi^+_c\to \Xi^0 \pi^+$ & 2.80 & 2.66 & 3.12 & 4.40 & 1.2$\pm$0.5$\pm$0.3\\
\hline
$\Xi^0_c\to \Lambda \bar K^0$ & 0.11 & 0.32 & 0.24 & 0.42 & \\
\hline
$\Xi^0_c\to \Sigma^0 \bar K^0$ & 1.03 & 0.08 & 0.12 & 0.20 & \\
\hline
$\Xi^0_c\to \Sigma^+ K^-$ & 0.11 & 0.11 &  & 0.27 & \\
\hline
$\Xi^0_c\to \Xi^0 \pi^0$ & 0.03 & 0.49 & 0.25 & 0.04 & \\
\hline
$\Xi^0_c\to \Xi^0 \eta$ & 0.21 & & & 0.28 & \\
\hline
$\Xi^0_c\to \Xi^0 \eta^\prime$ & 0.74 & & & 0.31 & \\
\hline
$\Xi^0_c\to \Xi^- \pi^+$ & 0.91 & 1.52 & 1.10 & 1.22 & \\
\hline
$\Omega^0_c\to \Xi^0 \bar K^0$ & 1.10 & & 0.08 & 0.02 & \\
\hline
$\Lambda^0_b\to \Lambda \pi^0$ & & & & 4.92$\times$10$^{-5}$ & \\
\hline
$\Lambda^0_b\to p K^-$ & & & & 2.11$\times$10$^{-4}$ & \\
\hline\hline
\end{tabular}
\end{center}

\vspace*{.5cm}
{\bf Table 2.} Branching ratios (in $\%$) for heavy-heavy transitions.
\begin{center}
\vspace*{.2cm}
\def\arraystretch{1}
\begin{tabular}{|c|c|c||c|c|c|}
\hline\hline
Process  & $\Gamma$ (in 10$^{10}$ s$^{-1}$) & $\alpha$  &
Process  & $\Gamma$ (in 10$^{10}$ s$^{-1}$) & $\alpha$  \\
\hline\hline
$\Lambda_b^0\to \Lambda_c^+\pi^-$     & 0.382 & -0.99 &
$\Xi_b^0\to\Xi_c^{\prime 0}\pi^0$ & 0.014 & 0.94\\
\hline
$\Lambda_b^0\to \Sigma_c^+\pi^-$      & 0.039 & 0.65 &
$\Xi_b^0\to\Xi_c^{\prime 0}\eta$ & 0.015 & -0.98\\
\hline
$\Lambda_b^0\to \Sigma_c^0\pi^0$      & 0.039 & 0.65 &
$\Xi_b^0\to\Xi_c^{\prime 0}\eta^\prime$& 0.021 & 0.97\\
\hline
$\Lambda_b^0\to\Sigma_c^0\eta$        & 0.023 & 0.79 &
$\Xi_b^0\to\Lambda_c^+ K^-$ & 0.010 & -0.73\\
\hline
$\Lambda_b^0\to\Sigma_c^0\eta^\prime$ & 0.029 & 0.99&
$\Xi_b^0\to\Sigma_c^+ K^-$ & 0.030 & -0.74\\
\hline
$\Lambda_b^0\to \Xi_c^0 K^0$          & 0.021 & -0.81 &
$\Xi_b^0\to\Sigma_c^0\bar K^0$ & 0.021 & 0\\
\hline
$\Lambda_b^0\to \Xi_c^{\prime 0} K^0$ & 0.032 & 0.98 &
$\Xi_b^0\to\Omega_c^0 K^0$ & 0.023 & 0.65\\
\hline
$\Xi_b^0\to \Xi_c^+\pi^-$ & 0.479 & -1.00 &
$\Xi_b^-\to \Xi_c^0 \pi^-$ & 0.645 & -0.97\\
\hline
$\Xi_b^0\to\Xi_c^{\prime +}\pi^-$& 0.018 & 0.61 &
$\Xi_b^-\to \Xi_c^{\prime 0}\pi^-$& 0.007 &-1.00\\
\hline
$\Xi_b^0\to \Xi_c^0\pi^0$ & 0.002 & -0.99 &
$\Xi_b^-\to \Sigma_c^0 K^-$ & 0.016 & -0.98\\
\hline
$\Xi_b^0\to\Xi_c^0\eta$ & 0.012 & -0.86 &
$\Omega_b^-\to \Omega_c^0\pi^-$ & 0.352 & 0.60\\
\hline
$\Xi_b^0\to\Xi_c^0\eta^\prime$ & 0.003 & 0.71 & & &\\
\hline\hline
\end{tabular}
\end{center}

\newpage
A clear pattern emerges. Nonfactorizing contributions plays a
significant role in the heavy-to-light transitions.
For heavy-to-heavy transitions the dominant
rates are into channels with factorizing contributions. The rates which
proceed only via nonfactorizing diagrams are small
but not negligibly small.

In Tables 3 and 4 we analyse the contributions of nonfactorizing diagrams
relative to those of the factorizing ones for the decays
$\Lambda_c^+\to\Lambda\pi^-$ and
$\Lambda_b^0\to\Lambda_c^+\pi^-$ which we predict to have the
largest branching ratio.
The total contribution of the nonfactorizing diagrams can be seen to be
destructive. The sum of nonfactorizing contributions amount
up to 60~\% of the factorizing contribution in amplitude of heavy-to-light
decay and up to 30~\% in amplitude of $b\to c$ transition.

\vspace*{.5cm}
\hspace*{2cm}{\bf Table 3.} Decay $\Lambda_c^+\to \Lambda\pi^+$.

\hspace*{2cm}Contribution of nonfactorizing diagrams

\hspace*{2cm}(in $\%$ relative to the factorizing contribution)

\begin{center}
\def\arraystretch{1.25}
\begin{tabular}{|c|c|c|c|c|}
\hline\hline
Amplitude   & \multicolumn{4}{|c|} {Diagram} \\
\cline{2-5}  & $II_a$  & $II_b$ & $II_a+II_b$ & $III$ \\
\hline\hline
A   & -29.8$\%$ & -18.5$\%$ & -48.3$\%$ & \\
\hline
B   & -32.4$\%$ & -15.9$\%$ & -48.3$\%$ &  -13.9$\%$  \\
\hline\hline
\end{tabular}
\end{center}

\vspace*{.5cm}
\hspace*{2cm}{\bf Table 4.} Decay $\Lambda_b^0\to \Lambda_c^+\pi^-$.

\hspace*{2cm}Contribution of nonfactorizing diagrams

\hspace*{2cm}(in $\%$ relative to the factorizing contribution)

\begin{center}
\def\arraystretch{1.25}
\begin{tabular}{|c|c|c|c|c|}
\hline\hline
Amplitude   & \multicolumn{4}{|c|} {Diagram} \\
\cline{2-5}  & $II_a$  & $II_b$ & $II_a+II_b$ & $III$ \\
\hline\hline
A   & -13.9$\%$ & -6.2$\%$ & -20.1$\%$ & \\
\hline
B   & -14.3$\%$ & -5.8$\%$ & -20.1$\%$ &  -8.5$\%$  \\
\hline\hline
\end{tabular}
\end{center}

\vspace*{1cm}
\hspace*{-1cm}
\unitlength=0.3mm
\special{em:linewidth 0.4pt}
\linethickness{0.4pt}
\begin{picture}(127.00,116.00)
\put(70.00,23.00){\oval(48.00,16.00)[]}
\put(45.00,23.00){\circle*{5.00}}
\put(95.00,23.00){\circle*{5.00}}
\put(95.00,23.00){\line(1,0){25.00}}
\put(120.00,24.00){\line(-1,0){25.00}}
\put(95.00,22.00){\line(1,0){25.00}}
\put(45.00,22.00){\line(-1,0){25.00}}
\put(20.00,23.00){\line(1,0){25.00}}
\put(45.00,24.00){\line(-1,0){25.00}}
\put(44.00,24.00){\line(3,5){26.00}}
\put(96.00,24.00){\line(-3,5){26.00}}
\put(70.00,67.00){\circle*{3.00}}
\put(13.50,23.00){\makebox(0,0)[cc]{\bf B}}
\put(128.00,23.00){\makebox(0,0)[cc]{\bf B$^\prime$}}
\put(70.00,78.00){\circle{14.00}}
\put(70.00,85.00){\circle*{3.00}}
\put(70.00,71.00){\circle*{3.00}}
\put(70.50,85.00){\line(0,1){15.00}}
\put(69.50,85.00){\line(0,1){15.00}}
\put(70.00,110.00){\makebox(0,0)[cc]{\bf M}}
\put(57.00,74.00){\makebox(0,0)[cc]{{\scriptsize\bf O$_\mu$}}}
\put(80.00,66.00){\makebox(0,0)[cc]{{\scriptsize\bf O$_\mu$}}}
\end{picture}
\vspace*{-4.7cm}
\begin{picture}(127.00,116.00)
\put(70.00,23.00){\oval(48.00,16.00)[]}
\put(46.00,23.00){\circle*{5.00}}
\put(94.00,23.00){\circle*{5.00}}
\put(95.00,23.00){\line(1,0){20.00}}
\put(95.00,24.00){\line(1,0){20.00}}
\put(95.00,22.00){\line(1,0){20.00}}
\put(45.00,23.00){\line(1,0){50.00}}
\put(45.00,22.00){\line(-1,0){20.00}}
\put(45.00,23.00){\line(-1,0){20.00}}
\put(45.00,24.00){\line(-1,0){20.00}}
\put(17.00,23.00){\makebox(0,0)[cc]{\bf B}}
\put(123.00,23.00){\makebox(0,0)[cc]{\bf B$^\prime$}}
\put(70.00,31.00){\circle*{3.00}}
\put(78.00,15.00){\circle*{3.00}}
\put(62.00,15.00){\circle*{3.00}}
\put(77.50,15.00){\line(0,-1){12.00}}
\put(78.50,15.00){\line(0,-1){12.00}}
\put(70.00,37.00){\makebox(0,0)[cc]{{\scriptsize\bf O$_\mu$}}}
\put(61.00,9.00){\makebox(0,0)[cc]{{\scriptsize\bf O$_\mu$}}}
\put(85.00,-5.00){\makebox(0,0)[cc]{\bf M}}
\end{picture}
\hspace*{-.3cm}
\begin{picture}(127.00,116.00)
\put(70.00,23.00){\oval(48.00,16.00)[]}
\put(46.00,23.00){\circle*{5.00}}
\put(94.00,23.00){\circle*{5.00}}
\put(95.00,23.00){\line(1,0){20.00}}
\put(95.00,24.00){\line(1,0){20.00}}
\put(95.00,22.00){\line(1,0){20.00}}
\put(45.00,23.00){\line(1,0){50.00}}
\put(45.00,22.00){\line(-1,0){20.00}}
\put(45.00,23.00){\line(-1,0){20.00}}
\put(45.00,24.00){\line(-1,0){20.00}}
\put(18.00,23.00){\makebox(0,0)[cc]{\bf B}}
\put(124.00,23.00){\makebox(0,0)[cc]{\bf B$^\prime$}}
\put(70.00,31.00){\circle*{3.00}}
\put(80.00,15.00){\circle*{3.00}}
\put(62.00,15.00){\circle*{3.00}}
\put(61.50,15.00){\line(0,-1){12.00}}
\put(62.50,15.00){\line(0,-1){12.00}}
\put(70.00,37.00){\makebox(0,0)[cc]{{\scriptsize\bf O$_\mu$}}}
\put(78.00,9.00){\makebox(0,0)[cc]{{\scriptsize\bf O$_\mu$}}}
\put(57.00,-5.00){\makebox(0,0)[cc]{\bf M}}
\end{picture}
\begin{picture}(127.00,116.00)
\put(70.00,23.00){\oval(48.00,16.00)[]}
\put(46.00,23.00){\circle*{5.00}}
\put(94.00,23.00){\circle*{5.00}}
\put(95.00,23.00){\line(1,0){22.00}}
\put(95.00,24.00){\line(1,0){22.00}}
\put(95.00,22.00){\line(1,0){22.00}}
\put(45.00,23.00){\line(1,0){50.00}}
\put(45.00,22.00){\line(-1,0){22.00}}
\put(45.00,23.00){\line(-1,0){22.00}}
\put(45.00,24.00){\line(-1,0){22.00}}
\put(15.00,23.00){\makebox(0,0)[cc]{\bf B}}
\put(127.00,23.00){\makebox(0,0)[cc]{\bf B$^\prime$}}
\put(70.00,31.00){\circle*{3.00}}
\put(70.00,23.00){\circle*{3.00}}
\put(70.00,15.00){\circle*{3.00}}
\put(70.50,15.00){\line(0,-1){12.00}}
\put(69.50,15.00){\line(0,-1){12.00}}
\put(75.00,35.00){\makebox(0,0)[cc]{{\scriptsize\bf O$_\mu$}}}
\put(61.00,26.50){\makebox(0,0)[cc]{{\scriptsize\bf O$_\mu$}}}
\put(80.00,-5.00){\makebox(0,0)[cc]{{\bf M}}}
\end{picture}

\vspace*{5.3cm}
\hspace*{1.25cm}I \hspace*{3.5cm}IIa\hspace*{3cm} IIb\hspace*{3.3cm} III

\vspace*{.5cm}
\begin{center}
{\bf Fig. 1 }
\end{center}

\newpage
{\Large\bf Acknowledgements}

\vspace*{.3cm}
\noindent
M.A.I, V.E.L and A.G.R thank Mainz University for the hospitality
where a part of this work was completed.
This work was supported in part by the Heisenberg-Landau Program,
by the Russian Fund of Basic Research (RFBR) under contract
96-02-17435-a, the State Committee of the Russian Federation for
Education (project N 95-0-6.3-67, Grant Center at S.-Petersburg State
University) and  by the BMBF (Germany) under contract 06MZ566.
J.G.K. acknowledges partial support
by the BMBF (Germany) under contract 06MZ566.

\vspace*{1cm}
{\Large\bf References}

\vspace*{.3cm}
\baselineskip 18pt
1. J.G. K\"{o}rner and M. Kr\"{a}mer, Z. Phys. {\bf C55} (1992) 659.

2. J.G. K\"{o}rner, in: Proc. VII Int. Conf. "Baryons '95" (Santa Fe, NM,

Oct. 1995), eds. P.D. Barnes et al. (World Sci., Singapore, 1996) p. 221.

3. Q.P. Xu and A.N. Kamal, Phys. Rev. {\bf D46} (1992) 270.

4. H.-Y. Cheng and B. Tseng, Phys. Rev. {\bf D48} (1993) 4188.

5. H.-Y. Cheng, Phys. Rev. {\bf D56} (1997) 2799.

6. M.A. Ivanov, J.G. K\"{o}rner, V.E. Lyubovitskij and A.G. Rusetsky,

Preprints MZ-TH/97-15, MZ-TH/97-21 (1997).

7. I.V. Anikin, M.A. Ivanov, N.B. Kulimanova and V.E. Lyubovitskij,

Phys. Atom. Nucl. {\bf 57} (1994) 1082; Z. Phys. {\bf C65} (1995) 681.

M.A. Ivanov and V.E. Lyubovitskij, Phys. Lett. {\bf B408} (1997) 435.

8. M.A. Ivanov, M.P. Locher, V.E. Lyubovitskij, Few-Body Syst. {\bf 21} (1996) 131.

9. M.A. Ivanov, V.E. Lyubovitskij, J.G. K\"{o}rner and P. Kroll,

Phys. Rev. {\bf D56} (1997) 348.

10. F. Hussain, J.G. K\"{o}rner and G. Thompson,
Ann. Phys. {\bf 206} (1991) 334.

11. F. Hussain, et al., Nucl. Phys. {\bf B370} (1992) 259.

12. Review of Particle Physics, Phys. Rev., {\bf D54} (1996) 1.
\end{document}